\documentclass{article}

     \PassOptionsToPackage{square, numbers}{natbib}

    \usepackage[preprint]{neurips_2020}

\usepackage[utf8]{inputenc} 
\usepackage[T1]{fontenc}    
\usepackage{hyperref}       
\usepackage{url}            
\usepackage{booktabs}       
\usepackage{amsfonts}       
\usepackage{nicefrac}       
\usepackage{microtype}      

\usepackage{graphicx}

\title{Feedback Effects in Repeat-Use Criminal Risk Assessments}

\author{%
  Benjamin D.~Laufer \\
  San Francisco, CA\\
  \texttt{ben.laufer@gmail.com} \\
}

\usepackage{amssymb}

\begin{document}

\maketitle

\begin{abstract}
In the criminal legal context, risk assessment algorithms are touted as data-driven, well-tested tools. Studies known as validation tests are typically cited by practitioners to show that a particular risk assessment algorithm has predictive accuracy, establishes legitimate differences between risk groups, and maintains some measure of group fairness in treatment. To establish these important goals, most tests use a one-shot, single-point measurement. Using a Polya Urn model, we explore the implication of feedback effects in sequential scoring-decision processes. We show through simulation that risk can propagate over sequential decisions in ways that are not captured by one-shot tests. For example, even a very small or undetectable level of bias in risk allocation can amplify over sequential risk-based decisions, leading to observable group differences after a number of decision iterations. Risk assessment tools operate in a highly complex and path-dependent process, fraught with historical inequity. We conclude from this study that these tools do not properly account for compounding effects, and require new approaches to development and auditing.
\end{abstract}

\section{Introduction}
As machine learning techniques have developed to replicate human decision-making, their use has forced a reconciliation with existing decision policies: can statistics do better? Are the statistics unfair, and are they more unfair than people?

A number of influential papers in 2015 \cite{kleinberg2015prediction, kleinberg2016inherent} suggested that accuracy in statistical forecasting methods can and should be used in ‘important’ contexts, where people’s freedom or health or finances are on the line, since these algorithms come with demonstrable accuracy levels. These contexts include sentencing and pre-trial decisions, credit scoring, medical testing and selective education access. Since then, the release of a ProPublica investigation of a common bail algorithm \cite{angwin2016machine} and retorts from the Criminology field \cite{dieterich2016compas, flores2016false} have forced a reckoning among theorists and practitioners about what fairness goals can and cannot be achieved.

Researchers have emphasized shifting focus from predictions to treatment effects, acknowledging that many of these high-impact decisions are, indeed, highly impactful on individual life-courses \cite{barabas2018interventions}. This revelation introduces the relatively new and under-analyzed topic of fairness in relation to repeated decision processes. Individual studies have demonstrated that ‘predictive feedback loops’ can lead to disproportionate over-policing in certain neighborhoods \cite{lum2016predict}, and that these loops can be modeled and simulated to demonstrate sub-optimal allocation in policing and compliance contexts \cite{ensign2018runaway, d2020fairness}.

The sequential-decision context is truly the norm, rather than the outlier. In virtually all high-impact scoring or testing systems, these processes occur (or may occur) numerous times throughout individual life-courses and each are both highly dependent on the past and highly impactful on individuals' futures. In light of sequential dependence in high-impact algorithms, this paper analyzes current methods for validating scoring systems as accurate and fair.

In the criminal legal context, new risk assessment algorithms are touted as data-driven, well-tested tools and often cite one or multiple validation studies that demonstrate a tool’s predictive accuracy and predictive parity between defendants of differing protected classes. Virtually all use a single-point-in-time, batch setting to analyze fairness and accountability concerns, with the exception of a few studies about how change in scores over time can better predict future scores \cite{skeem2016risk, labrecque2014importance, vose2016risk, latessa2016does}. We show that these tests are not catered to the criminal legal domains, where decisions often occur sequentially at multiple times through a defendant's life. We take a close look at the statistical methods used by these studies, and show using simulation experiments that risk assessment tests can fail at meeting a number of fairness definitions even while passing instantial validity tests.

\subsection{Validation and One-Shot Testing}

Risk assessment algorithms are developed and then tested for `validity’. These experiments, formerly only concerned with predictive validity, now test various potential biases that algorithms may exhibit in new populations. Validation experiments have therefore become an important aspect of the risk-assessment development process, and validity is seen as a necessary requisite for any risk assessment algorithm in use. What does validity mean?

While there has been some controversy over the way in which risk assessment tools get developed,\footnote{In Philadelphia, for example,  recidivism was being measured as re-arrest rate, and because of public opposition the sentencing commission began measuring it as subsequent conviction rate.} remarkably little analysis has been conducted of the best practices for validation in risk assessment. As a result, many validation experiments resemble one another. Typically, the studies measure a tool’s predictive capacity by analyzing post-conviction arrest rates over a short time-frame. They take a group of defendants released from the same jurisdiction in a given time-frame, and determine the average re-arrest rate of defendants with different risk scores over a typical period of one or two years. For example, Lowenkamp et al. conducted a validation experiment in which they tested the LSI-R and the LSI-Screening Version, which screens defendants to decide whether to administer the more in-depth LSI-R assessment \cite{lowenkamp2009validating}. Using a look-ahead period of 1.5 years, the study measured re-arrest rate and re-conviction rate, and found that a higher LSI-R score is positively correlated with future incarceration.

Interestingly, algorithmic risk assessments tend to find disparate validity levels when the same algorithm is used on racially distinct populations. Fass et al. in 2008 published validation data on the Level of Service Inventory - Revised (LSI-R) algorithm, as well as COMPAS \cite{fass2008lsi}. Using a dataset of 975 offenders released into the community between 1999-2002 from New Jersey, the measurement period was 12 months. The purpose of the study was to see whether these algorithms, trained on mostly white populations, are invalid for a population like New Jersey, which has has ``substantial minority” representation in incarceration. The study finds ``inconsistent validity when tested on ethnic/racial populations” \cite[1095]{fass2008lsi}, meaning the predictive validity may suffer as the result of differences between the training cohort used to develop the algorithm and the actual demographic breakdown of a jurisdiction. Demichele et al. in ``The Public Safety Assessment: A Re-Validation” use data from Kentucky provided by the Laurence and John Arnold Foundation, which developed the PSA. The study measured actual failure-to-appear, new criminal activity, and new violent criminal activity before a trial. They found that the PSA exhibited broad validity, but found a discrepancy based on race \cite{demichele2018public}.

Beyond recidivism, a few studies have focused on the relationship between risk assessment-driven decisions and other life outcomes, including earnings and family life. Bruce Western and Sara McLanahan in 2000 published a study entitled ``Fathers Behind Bars” that finds alarming impacts of incarceration on family life. A sentence to incarceration was found to lower the odds of parents living together by 50-70\% \cite{western2000fathers}. Dobbie et al. published a study that demonstrated that pre-trial detention in Philadelphia on increased conviction rates, decreased future income projects and decreased the probability that defendants would receive government welfare benefits later in life \cite{dobbie2018effects}. The Prison Policy Initiative reports an unemployment rate above 27\% for formerly incarcerated people, and find a particularly pronounced effects of incarceration on employment prospects for women of color \cite{ppi}. 

Given the deeply impactful nature of risk-based decisions, validation experiments are surprisingly limited in scope. The outcome variable - typically rearrests in a one or two-year window - fail to capture the many ways that a risk-assessment can impact an individual’s family, employment, income, and attitudes - all of which may be relevant in considering recidivism. Perhaps more importantly, the various aspects of life impacted by detention are precisely the risk factors that may get picked up by a subsequent judicial decision.

By treating risk assessment as instantial and analyzing longitudinal effects of a single assignment of risk, validation experiments are only observing part of the picture. When we consider the tangible impacts of judicial decisions and relate these impacts to future decisions, we see that there are possible feedback effects in the criminal system. The dependence of subsequent judicial decisions on prior judicial decisions is rampant. Sentencing guidelines suggest (and often require) judges to give longer sentences to repeat offenders, for example. The very notion of responsivity in criminal treatment requires periodic assessments that determine the `progress’ or treatment effect over time for a given defender, and shape punishment accordingly. However, treatment of sequential risk-assessments and the possible harms of feedback is missing from a literature that has so exhaustively debated whether incarceration has a criminogenic effect.

This paper explores how compounding in criminal justice impacts defendants. The treatment of risk assessment as innocuous, objective, statistical prediction has clouded rigorous theoretical exploration of lifetime compounding in criminal punishment. Using data from Philadelphia, we find that higher confinement sentences significantly increase cumulative future incarceration sentences for defendants. Synthesizing data from Philadelphia with a theoretical understanding of feedback in algorithmic risk assessment, we will discuss implications for judges and defendants.

\subsection{Contributions}

This paper is meant to critically evaluate the current vetting and auditing process for high-stakes, repeated-use risk assessment algorithms that are deployed in the U.S. criminal legal system.

First, we develop a generalized sequential scoring-decision model, which can be used in simulation experiments to test for possible compounding effects in group fairness, uncertainty, and punishment. Then, using simulation experiments, we demonstrate that a risk assessment can pass validity tests and still exhibit problems with predictive accuracy, group-fairness, and risk-group-difference.

The broader argument put forward by this paper is that current validation tests do not consider sequential feedback, and are therefore insufficient to approve criminal risk assessments for use. Algorithms used in the criminal legal system, credit system, and in other high-impact domains should test for unintended impacts when used repeatedly.

\section{Model Problem Setting}

We offer a model of repeated high-impact decisions that will help us simulate the purpose and pitfalls of validation tests. We use a binary observation-decision system that allows each decision to impact the underlying propensity for a failed observation.

We can imagine this context as being a repeated parole decision, where an officer uses a risk score at each meeting to decide whether to impose a more restrictive policy on a parolee (e.g. curfew), thus limiting employment opportunities and increasing the probability of unlawful behavior. Each periodic parole meeting there is some observation of whether the rules were broken, a re-assessment of risk, and a new binary treatment decision. The context also has parallels in credit decisions, regulatory compliance checks, ad clicks, and more.

\subsection{General Modelling Assumptions}

We begin with a simple model of risk-needs driven decisions. Given that existing risk assessment services emphasize their wide applicability, some algorithms are adopted at numerous stages in criminal proceedings. Other jurisdictions may use different assessments for policing, bail, sentencing and parole. Starting simple, we model risk assessments as instantaneous binary decisions that are separated in time. Each decision occurs sequentially, and the outcome is either “high risk” or “low risk”, as visualized in Figure 1.

\begin{figure}[ht]
    \centering
    \includegraphics[width=0.8\linewidth]{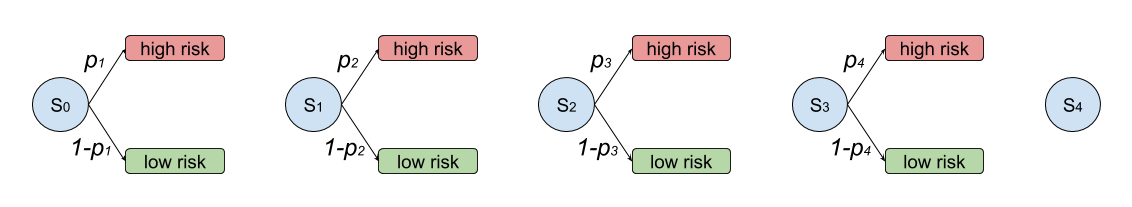}
    \caption{Sequential decision context diagram}
    \label{fig:my_label}
\end{figure}

We assume here that risk assessments are conducted $T$ times throughout a person’s life, and that the assessment $r_t$ measures some underlying probability of future criminality $p_t \in [0,1]$. The risk assessment $r$ fully dictates a decision $X_i$, which denotes some choice of high-risk or low-risk treatment (e.g. increased surveillance, or prison security level):
$$X_i \in \bigg{\{} \begin{array}{c}
    \ 1, \ \ \ \ \ \textit{if defendant is classified high-risk} \\
    0,\ \ \ \ \ \textit{if defendant is classified low-risk}
\end{array}$$
We model each assessment using the current state of the world before decision $i$, denoted $S_{i-1}$.

The assessment is a random variable and not deterministic because risk assessment algorithms do not solely determine defendant outcomes - the ultimate decision is still up to a judge, who references the risk assessment score as part of the broader pre-trial policy decision.

We wish to explore the possibility that outcomes of assessments may impact and alter future assessments. As such, our model must enable us to analyze cases where the outcome variable $X_i$ may impact the probability of high-risk classification for $X_{i+1}, X_{i+2}, ... , X_N$. The probability of a high-risk classification at decision $i$ can thus be thought of as a function of some defendant information $D_i$ (gender, race, age) and the history prior decisions, $H_i$. We write the current state of beliefs at $i$ as $S_i = \{D_i, H_i\}$. We more accurately portray this dependence on the history of decisions as a branching process, rather than a sequence of decisions, in Figure 2.

\noindent
\begin{figure}[ht]
    \centering
    \includegraphics[width=0.5\linewidth]{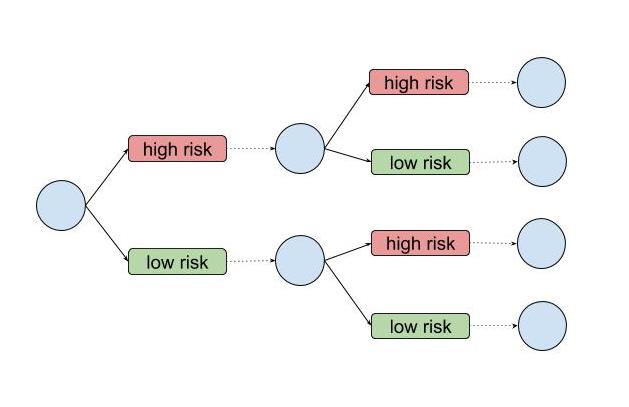}
    \caption{Branching and Path Dependence in a Binary Risk Classification Scorer}
    \label{fig:my_label}
\end{figure}

Every major risk assessment algorithm uses information about criminal history to assess risk. PSA, for example, measures a defendant's number of prior misdemeanors, felonies, convictions, and violent convictions. These numbers add various point values to a risk assessment score, and a threshold value may determine pre-trial detention or cash bail amounts. Therefore, the PSA and most (if not all) other algorithms have a reinforcement effect. After an individual is convicted with a felony charge, every subsequent risk assessment for the rest of his life will use his criminal history to increase his risk score. Thus, initial assessments of risk can hold more `weight' in determining lifetime treatment than later assessments. If a person is identified as high-risk in their first encounter with the criminal system, known effects on future crime rates, employment, family life, taxes, and other features will increase the likelihood of subsequent encounters.

This property of \textit{reinforcement} is key to modeling our system. The process is not Markovian: history matters, and our state of beliefs changes over time. Instead, we understand the changing effects of sequential risk-assessments as an Urn process, derived from the classic Pólya Urn model in mathematics \cite{pemantle2007survey}.

\subsubsection{Dependence and Reinforcement}

\ 

\noindent
Let's say each risk assessment decision affects subsequent decisions as follows: If $X_{i-1}$ is the risk-assessment outcome for decision $i-1$, the subsequent probability of a high-risk decision $p_{i}$ is a weighted average between $p_{i-1}$, the prior probability, and $X_{i-1}$, the most recent classification:
$$p_{i} = p_{i-1}\left[\gamma_{i}\right] + X_{i-1}\left[1-\gamma_{i}\right], \ \ \ \ i \in \{2,...,N\},\ \ \gamma_i \in [0,1]$$
This means that we model updates in risk score by averaging the prior assumed risk and the outcome of a new assessment. The $X_{i-1}$ term can be thought of as the marginal effect of a new classification on defendant risk.  To model reinforcement, we allow $\gamma_i$ to increase as $i$ increases, letting prior risk score $p_{i-1}$ hold more importance as a defendant is older and has more history. This should make intuitive sense - if a defendant has lived out most of his life with a certain propensity for criminal activity (`risk'), the effect of a new assessment should carry less weight.

Using the above intuition, we'll start by assuming the following relationship between $\gamma_i$ and $i$ (the number of encounters with the criminal justice system): 
$$\gamma_i = \frac{i}{i+1}$$
To understand the equation above, let's consider the value of $\gamma_i$ for varying $i$. In a first encounter with criminal courts where $i=1$, we'd have $\gamma_1 = \frac{1}{2}$. Risk assessment outcome $X_1$ would thus have a very strong impact on future risk assessments. When $i$ is high, however, $\gamma_i$ approaches $1$ and new assessments would diminish in weight. This is the reinforcement property we're seeking - the more decisions that go by, the less weighty they are in determining a person's lifetime experience with the state's criminal system.

Thus, our formula for $P(X_{i} | D,H_i)$ is:
\begin{equation} \label{defendant_eq}
P(X_{i} | p_{i-1}, X_{i-1}) = p_{i-1}\left[\frac{i}{i+1}\right] + X_{i-1}\left[\frac{1}{i+1}\right], \ \ \ \ i \in \{2,...,N\}
\end{equation}
Let's assume temporarily that every defendant starts off with a probability of high-risk classification $p_1=\frac{1}{2}$. We model the effect of sequential risk-assessments for different defendants by implementing our iterative equation. Below are sample paths for 5 defendants who are subject to ten periodic, evenly spaced assessments over time:

\noindent
\begin{center}
    \includegraphics[scale = 0.38]{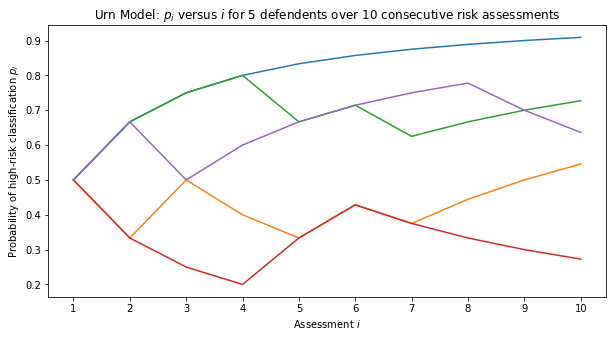}
\end{center}

In the plot above, each color represents an individual who encounters criminal risk assessments throughout their life. Notice that this plot behaves in accordance with the reinforcement effect - initial assessments have large effects on $p_i$, and later assessments only marginally change the course of the risk level. Indeed, the for very large $i$ the risk level approaches a straight-line, meaning that the system reaches a stable propensity for criminal activity. Below are the paths of the same five defendants, this time over a total of 100 assessments (so 90 additional assessments):
\begin{center}
    \includegraphics[scale = 0.38]{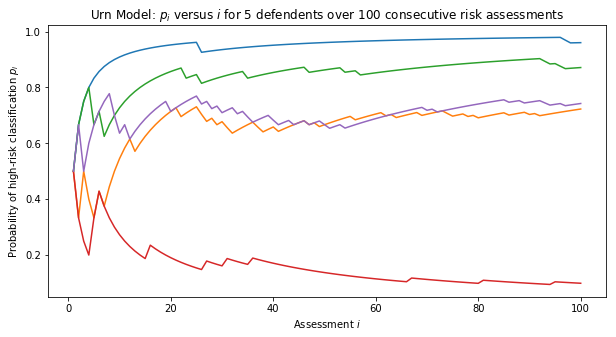}
\end{center}

While it is unrealistic that a single person would have one hundred exactly evenly spaced and identical assessments throughout their lives, the behavior of our model seems to cohere with our knowledge of risk-assessments - their output impacts future assessments in a way that reinforces their classification. In other words, people detained after being identified as high-risk are more likely to re-offend, spend time in jail, have financial trouble, lose employment, or receive a guilty charge - all of which will affect their level of `risk'.

\subsubsection{Pòlya's Urn Generalization}

\ 

\noindent
The model derived above is an Urn process. Borrowing a few theorems from probability theory, we can begin to understand the large-scale, long-term effects that might come about when algorithms are used consecutively throughout a person's life.

Pòlya's Urn can be used to model path-dependent branching processes that are 'exchangeable', meaning the order of prior events does not matter.\footnote{This is an assumption that may not hold true for our case, because many algorithms care about how \textit{recent} a historical event took place. PSA, for example, cares about prior failures to appear in court in the past two years. However, for the most part, algorithms consider the aggregate number of historical events - number of prior felonies, misdemeanors, convictions, etc. These indicators are all \textit{exchangeable} in the sense that it doesn't matter when in the defendant's life they occurred.} The model asks what the long-term distribution of blue balls will be in the following random process:

\ 

\begin{itemize}
    \item An urn contains $R_t$ red balls and $B_t$ blue balls. Start at $t=0$, with an initial mix of $R_0$ and $B_0$ balls.
    \item for iteration $t \in \{1,...,T\}$:
    \begin{itemize}
        \item Pick a ball randomly from the urn.
        \item For the ball picked, return it and $k$ additional balls of the same color to the urn.
    \end{itemize}
\end{itemize}

\subsubsection{Urn Equivalence to a Risk Assessment Model}

\ 

\noindent
We can model reinforcement in algorithmic decision-making as an urn process. Our basic defendant model replicates exactly the basic Pòlya process with $R_0 = 1$, $B_0 = 1$, and $k=1$. We derive the equivalence in the two processes below.

\noindent
Denote the color of the ball selected by pick $i \in \{1,2,...,N\}$ as:
$$\tilde{X}_i \in \bigg{\{} \begin{array}{c}
    \ 1, \ \ \ \ \ \textit{if blue ball is picked} \\
    0,\ \ \ \ \ \textit{if red ball is picked}
\end{array}$$

\noindent
Assuming each ball is picked with equal probability, the probability of picking blue in is given by:
$$P(\tilde{X}_i = 1) = \frac{B_{i-1}}{B_{i-1} + R_{i-1}}$$

The total number of ball in the urn is $n_i = R_i + B_i$. The probability of picking blue  given all prior picks is denoted as $\tilde{p}_i$. We can always find $\tilde{p}_i$ by dividing the number of blue balls in the urn by the total number of balls. We've shown that $p_i = \frac{B_{i-1}}{n_{i-1}}$. After the $i^{th}$ pick, what will be the probability of picking blue? We inevitably add $k$ balls into the urn, so $n_{i} = n_{i-1} + k$. In the event that our pick is red, we still have $B_{i-1}$ blue balls, so the probability of picking blue decreases to $\frac{B_{i-1}}{n_{i-1} + k}$. If we do pick blue, however, the probability increases to $\frac{B_{i-1} + k}{n_{i-1} + k}$. Thus, the probability of picking blue on the $(i+1)^{th}$ pick, given $B_0, n_0$ and $\tilde{X}_1$, is:
$$\tilde{p}_{i+1} = \frac{B_{i-1} + \tilde{X}_ik}{n_{i-1} + k} $$

With a bit of algebra, we can define this probability in terms of the probability for the prior pick:
$$\tilde{p}_{i+1} = \frac{B_{i-1}}{n_{i-1} + k} + \tilde{X}_i\frac{k}{n_{i-1} + k} = \left[\frac{B_{i-1}}{n_{i-1}}\right]\frac{n_{i-1}}{n_{i-1}+k} + \tilde{X}_i\frac{k}{n_{i-1} + k}$$
$$\therefore\  \tilde{p}_{i+1} = \tilde{p}_{i}\frac{n_{i-1}}{n_{i-1}+k} + \tilde{X}_i\frac{k}{n_{i-1} + k}$$

When $k=1$ and $R_0 = B_0 = 1$, how does $n_i$ behave? It starts at $n_0 = 2$, and after each pick it increments by $k=1$. Thus, $n_i = 2+i$. Equivalently, $n_{i-1} = 1+i$, and $n_{i-2} = i$. Using the relationship derived above, a shift in index yields the probability of picking blue $\tilde{p}_i$ for $i \in \{2,...,N\}$:
\begin{equation} \label{equation2}
\tilde{p}_{i} = \tilde{p}_{i-1}\frac{n_{i-2}}{n_{i-2}+k} + \tilde{X}_{i-1}\frac{k}{n_{i-2} + k} = \tilde{p}_{i-1}\left[\frac{i}{i+1}\right] + \tilde{X}_{i-1}\left[\frac{1}{i+1}\right]
\end{equation}

Notice the equivalence to equation \ref{defendant_eq}. We've shown the probability for picking blue at each iteration of the classic Pólya Urn process exactly equals the probability of a high-risk classification in our simple model of sequential risk assessments, where $\tilde{p}_i = p_i$ and $ \tilde{X}_i = X_i$.

\subsection{Long Run Behavior}
When we say that a sequence of random decisions might exhibit \textit{reinforcement}, we now know that this means something deeper mathematically. Random processes with reinforcement behave in certain ways that might be problematic in the context of criminal policy. We have a general sense that algorithmic decisions in criminal justice impact defendants profoundly, and likely impact future encounters with law enforcement. Leveraging insights from probability theory, we can begin to understand the danger of policies that have compounding effects.

To start, we analyze the long-term treatment of individuals that are subject to sequential risk-based decisions. In Robin Pemantle's ``A Survey of Random Processes with Reinforcement" (2006), the following theorem is reported about Pòlya's Urn process:

\begin{quote}
    Theorem 2.1: The random variable $p_i = \frac{B_i}{B_i + R_i}$ converges almost surely for large $i$ to a limit $P$. The distribution of $P$ is: $P \sim \beta (a,b)$ where $a = \frac{B_0}{k}$ and $b = \frac{R_0}{k}$. In the case where $a=b=1$, the limit variable $P$ is uniform on $[0,1]$. \cite{pemantle2007survey}
\end{quote}

Theorem 2.1 lays out how we can expect our modeled risk assessments to behave over many iterations. If one person undergoes risk assessments numerous times throughout their life, they may end up in radically different places depending on the risk-assessment outcome. They may be able to steer clear of subsequent confinement and re-arrest, or they may be continuously surveiled and repeatedly penalized by the state. 

For a preliminary understanding of how inter-dependence in repeated risk assessments can impact a population, we use our initial modeling assumption that $p_1 = 0.5$ (so $B_0 = R_0$ and $a = b$), and imagine varying the parameter that determines the bearing of prior assessments on updated assessments, $k$ (which defines $\gamma$). If we decrease $k$ to $0.1$ so that $a = b = \frac{B_0}{k} = 10$, we have the following long-term distribution for defendant risk. See Figures 3 and 4.

\noindent
\begin{figure}[ht]
    \centering
    \caption{PDF of long term risk level when $k = 0.1$}
    \includegraphics[scale = 0.3]{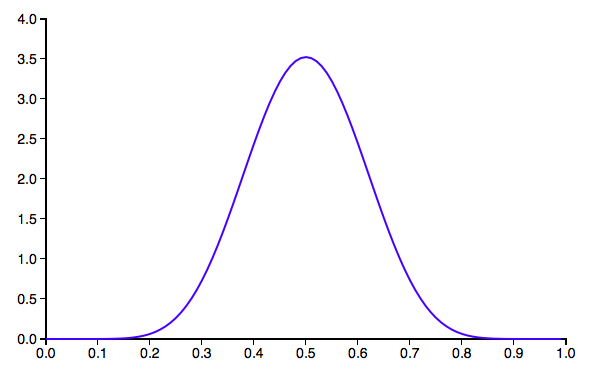}
    \label{fig:my_label}

    \caption{Urn Model Plot, $p_i$ versus $i$ for 30 defendants over 15 consecutive risk assessments, $k=0.1$}
    \includegraphics[scale = 0.45]{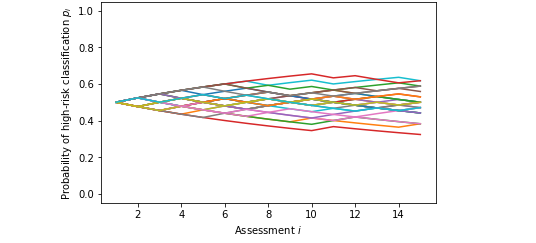}
\end{figure}

When decisions have little impact on people's lives (and potential subsequent risk assessments), we see consistency in long-term outcomes. Everyone starts with a risk score of $0.5$, and all end up somewhere near there even after many assessments.

However, if algorithmic-driven decisions are more sensitive to the effect of prior decisions with $a = b = \frac{B_0}{k} = 0.1$, then we can see very problematic behavior in the long term. See Figures 5 and 6.

\noindent
\begin{figure}[ht]
    \centering
    \caption{PDF of long term risk level when $k = 10$}
    \includegraphics[scale = 0.3]{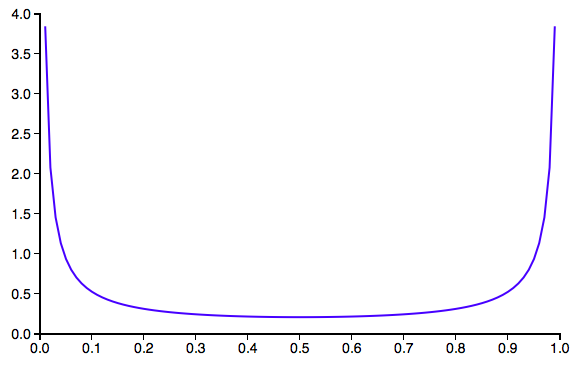}

        \caption{Urn Model Plot, $p_i$ versus $i$ for 30 defendants over 15 consecutive risk assessments, $k=10$}
    \includegraphics[scale = 0.45]{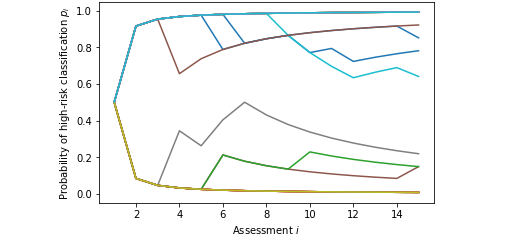}
    \label{fig:my_label}
\end{figure}

In this second case, we begin with defendants that are identical in attributes, with an initial probability of high-risk classification $p_1 = 0.5$. However, simply because of the effect of risk-based decision making, defendants end up with radically different risk levels, and are highly likely to be pushed to an extreme (no criminal risk, 0, and extreme criminal risk, 1).

Of course, these results are purely theoretical and do not come from real observed processes. But they motivate the importance of scrutinizing how algorithms are used in practice. Algorithms may be validated to ensure that biases are mitigated to a certain confidence threshold. But even tiny disparities in the system described by the second plot above can profoundly impact outcomes.

\section{Discussion}

Understanding that sequential feedback-effects exist in criminal legal decisions forces us to re-evaluate the ways that validations are currently used. 

The effect of prison time and similar decisions on future encounters with criminal punishment implies that algorithmic risk-assessment tools cannot be assessed using instantial experiments at one time in a defendant’s life. If larger sentences are associated with greater prison time, it is likely that longer sentences hold bearing on future risk assessment. A more severe sentence may lead parole officers to have more discretion over parolees. It may increase a defendant’s association with other criminals. This kind of dependence between decisions is clear from sentencing tables and three-strikes rules, which recommend that judges give exaggerated sentences to repeat-offenders.

Since judicial decisions appear to feed into one another sequentially over a defendant’s life time, it is important to consider models that encompass compounding effects. Risk assessment algorithms and validation experiments fail to adequately address the potential of feedback effects over time. Rigorously considering the impacts of dependent, sequential decisions will be necessary for deploying any high-stakes algorithm.

\section*{Broader Impact}

My hope is that this inquiry exposes some of the shortcomings of auditing in high-impact ML domains. The discussion and analysis were specifically about the criminal legal space; however, many of the findings are relevant to the use of high-impact ML algorithms in many fields. In credit and medicine, for instance, risk determinations are premised on historical access to resources (e.g. capital or medical attention), so when future triage decisions are made, risk-based decisions will always exhibit the effects of historical decisions. None of these systems should treat risk as exogenous or innate and should instead have the goal of \textit{minimizing harm}.

\section*{Acknowledgments}

I'd like to acknowledge Miklos Racz, my undergraduate research advisor who has been helping me pursue and build on my research after graduating.

I'd like to acknowledge my friends, family, colleagues, and role models who have provided me with all the skills and access necessary to submit to a venue like this one.

\small

\bibliography{my_bibliography}

\end{document}